\documentclass[aoas,MSNbibl,nameyear,dvips]{arximspdf}
\usepackage{mathbh}
\usepackage{dcolumn}
\usepackage{graphicx}

\doi{10.1214/11-AOAS513}
\volume{6}
\issue{1}
\pubyear{2012}
\firstpage{106}
\lastpage{124}

\makeatletter

\def\bsuffix #1{#1}

\newcolumntype{d}[1]{D{.}{.}{#1}}

\newcommand{\bbE}{\mathbb{E}}
\newcommand{\bbR}{\mathbb{R}}
\newcommand{\mcH}{\mathcal{H}}
\newcommand{\mcT}{\mathcal{T}}

\makeatother

\begin{document}
\begin{frontmatter}

\title{Self-exciting hurdle models for terrorist activity}
\runtitle{Self-exciting hurdle models for terrorist activity}

\begin{aug}
\author[A]{\fnms{Michael D.} \snm{Porter}\corref{}\ead[label=e1]{porter.mike@geoeye.com}}
\and
\author[B]{\fnms{Gentry} \snm{White}\ead[label=e2]{gentry.white@uq.edu.au}}
\runauthor{M. D. Porter and G. White}
\affiliation{GeoEye Analytics and University of Queensland}
\address[A]{GeoEye Analytics\\
7921 Jones Branch Drive, Suite 600\\
McLean, Virginia\\
USA\\
\printead{e1}}
\address[B]{Institute for Social Science Research\\
ARC Centre of Excellence in Policing\\
\quad and Security\\
University of Queensland\\
St. Lucia, Queensland 4072\\
Australia\\
\printead{e2}}
\end{aug}

\received{\smonth{6} \syear{2011}}
\revised{\smonth{9} \syear{2011}}

%
\begin{abstract}
A predictive model of terrorist activity is developed by examining the
daily number of terrorist attacks in Indonesia from 1994 through 2007.
The dynamic model employs a shot noise process to explain the
self-exciting nature of the terrorist activities. This estimates the
probability of future attacks as a function of the times since the past
attacks. In addition, the excess of nonattack days coupled with the
presence of multiple coordinated attacks on the same day compelled the
use of hurdle models to jointly model the probability of an attack day
and corresponding number of attacks. A power law distribution with a
shot noise driven parameter best modeled the number of attacks on an
attack day. Interpretation of the model parameters is discussed and
predictive performance of the models is evaluated.
\end{abstract}

%

\end{frontmatter}

\section{Introduction}

Since the attacks of September 11, 2001 there has been a substantial
increase in the amount of resources
dedicated to combating terrorism.
As a single example, 
the Congressional Research Services estimates that
as of September 2010 \$1.2 trillion USD has
been spent on the Global War on Terror, including \$28.5
billion USD on enhanced security [\citet{Belasco2010}]. Despite
this increase in spending, there are no accurate measures of the
effectiveness of these counter-terrorism
efforts [\citet{Lum2006}, \citet{Perl2007}]. According to the
Congressional
Research Service, one of the keys to combating
terrorism effectively is to understand, and have correctly specified
models for,
terrorist activity:

\begin{quote}
``Better
understanding of the dynamics of terrorism allows for a more complete
picture of
the complexities involved in measuring success or failure and can
assist the
110th Congress as it coordinates, funds, and oversees anti-terrorism
policy and
programs'' [\citet{Perl2007}].
\end{quote}
%
A dynamic model for terrorism
should describe terrorist activity in a way that
is consistent with both the theoretical framework for
terrorism and the observed data.\vadjust{\goodbreak} 
Defining accurate models of terrorist activity
is important not just from the standpoint of assessing the
effectiveness of counter-terrorism efforts but also for use as a tool to
predict the future risk of terrorist attack.

A majority of publications in the literature use one of three basic modeling
approaches to terrorism. The work of
Enders and Sandler
(\citeyear
{EndersSandler1993,EndersSandler2000,EndersSandler2002,EndersSandler2006})
and \citet{Barros2003} uses
time series
analysis techniques including intervention \mbox{analysis} and vector autoregressive
(VAR) models. Other more recent papers have applied group based trajectory
analysis [\citet{Nagin2005}] to analyze terrorist data. Examples
of this
include \citet{LaFreeetal2010} who incorporate a zero-inflated Poisson
distribution to account for an excess of zero counts.
Other approaches use Cox proportional hazards models to directly
model the time between attacks
[\citet{Duganetal2005}, \citet{LaFreeetal2009}]. These last two
papers attempt to account for the recent event
history as influencing current events by constructing an ad hoc term
that describes
the recent density of attacks.
While this is a step toward incorporating the dynamics related to the
process history, it requires prespecification of the form of dependence.

The desire to include terms to account for recent event history is
rooted in
the theoretical understanding of terrorism and other politically motivated
violence.
The clustering hypothesis,
concurrent with contagion theory [\citet{Midlarsky1978}], explains
politically
motivated violent activity as a series of nonindependent events, where each
event influences the probability of subsequent events. This hypothesis
is well
accepted in criminological and sociological approaches to
terrorism. Clustering effects and contagious behavior have been
demonstrated in military coups [\citet{LiThompson1975}], international
terrorism
[\citet{Midlarskyetal1980}], airline hijackings [\citet
{Holden1986}], race
riots [\citet{Myers2000}] and insurgent activity [\citet
{Townsleyetal2008}].

In examining terrorism data in general there are two major issues to address.
The first is that as terrorist attacks are usually rare, the daily
number of terrorist attacks are often zero, but due to large
coordinated attacks there are some extreme values.
This combination of a large number of zeros and extreme values is poorly
modeled by standard probability distributions.
Second, the timing of terrorist incidents appear to be clustered,
rendering standard models that assume independence unsuitable.

The first of these issues is addressed 
using a hurdle model
[\citet{Mullahy1986}], also known as the two-part model
[\citet{Heilbron1994}].
The hurdle model is a two component model that allows separate
specifications of the probability of a zero count and the probability
of a nonzero count. This allows the hurdle model to accommodate a large
number of zeros in addition to some extreme counts.
The hurdle model is used in a variety of applications, including in
ecology for
modeling counts of rare species [\citet{Welshetal1996}], in public
health for
modeling smoking behavior [\citet{Jones1994}],
in political science for modeling proportional representation in minority
electorates [\citet{Marschalletal2010}] and\vadjust{\goodbreak}
network change detection
[\citet{Heardetal2010}].
For terrorism modeling, the hurdle model
is preferred to the zero-inflated model
[\citet{Lambert1992}] which assumes that the extra
zeros observed are due to censoring. The hurdle model assumes that the extra
zeros are due to a separate process (the ``hurdle''), which must be
overcome before the number of corresponding incidents are determined.
This is more reasonable for terrorism data, as it can often be assumed
that sparsity of attacks is because they are indeed rare, not because
they are unobserved.

The second issue regarding the clustering behavior of terrorist
activity is addressed by incorporating a self-exciting component
[\citet{Hawkes1971a}]. The self-exciting component specifies that the
probability of a event is a function of the time (and possibly other
aspects) of all previous events, such that the effect on the
probability decreases over time. This can help account for the
clustering and dynamic nature of terrorism. For example,
Holden (\citeyear{Holden1986,Holden1987}) developed a self-exciting
Poisson model
that better represented the dynamics of airline hijacking and
\citet{Mohleretal2010} a self-exciting space--time model for residential
burglaries.

This paper combines these two concepts to examine the use of
self-exciting hurdle models for terrorist activity in Indonesia and
Timor-Leste between 1994 and 2007. The hurdle component is modeled as a
self-exciting Bernoulli process where the form of self-excitation is
dictated by the data.
The nonzero counts are modeled by an extreme value distribution with
parameters that are also a function of the event history.
The corresponding self-exciting hurdle model is capable of making
predictions and providing information about the risk of terrorist
activity without any additional covariate information other than the
event history.
This capability is important as covariate information for terrorism
data is often either missing or unreliable.

\section{Data and exploratory analysis}
\label{data}

The Global
Terrorism Database [La\-Free and Dugan (\citeyear{LaFreeDugan2007})] is an open-source publicly
available
database of over 87,000 terrorist events around the world from 1970 through
2008. The
database is continually updated with new information and is believed to
be the
most comprehensive database of its kind. The data used here are a
subset of the
GTD consisting of daily
counts of terrorist attacks in Indonesia\footnote{This includes data
from Timor-Leste (East Timor) which became a sovereign state in 2002.}
from January 1, 1994 through December~31, 2007.

\subsection{Exploratory data analysis}

\begin{figure}

\includegraphics{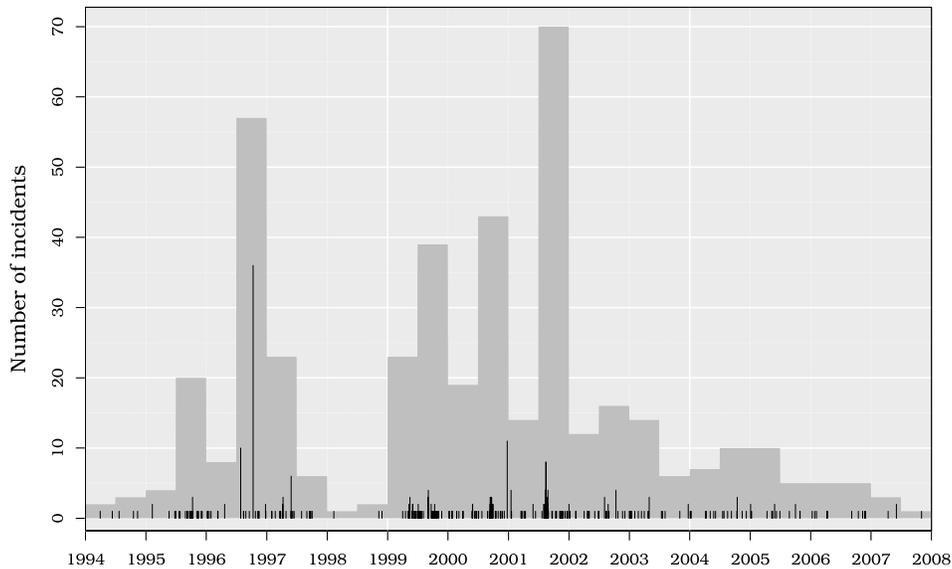}

\caption{Daily number of terrorist events in Indonesia. The histogram
shows the cumulative number of attacks in each six month period.}
\label{figdata}
\end{figure}

The data from a training period of 1994 through 2000 was examined to
inform model construction.
Of the 2,557 days considered in this analysis, there were 250 terrorist
attacks on 158 unique event days.
Figure \ref{figdata}, displaying the daily and semi-yearly counts of
attacks, illustrates the nature of the terrorist activity.
In particular, the attacks appear clustered\vadjust{\goodbreak} with long stretches between
attacks followed by a period of increased activity. In addition, while
most days have no attacks (93.8\%), some days have multiple attacks
(possibly due to planned coordinated attacks).\looseness=-1

%
%
\begin{table}[b]
\caption{Distribution of observed and expected number of Indonesian
terrorist attacks per day (1994--2000)} 
\label{distribution}
\begin{tabular*}{\tablewidth}{@{\extracolsep{\fill}}ld{4.0}d{4.1}d{4.1}@{}}
\hline
\multicolumn{1}{@{}l}{\textbf{\# Attacks}}
& \multicolumn{1}{c}{\textbf{\# Days}} &
\multicolumn{1}{c}{\textbf{Poisson}}
& \multicolumn{1}{c@{}}{\textbf{Neg.Bin}} \\
\hline
\hphantom{$>$}0 & 2\mbox{,}399 & 2\mbox{,}319 & 2\mbox{,}401 \\
\hphantom{$>$}1 & 130 & 227 & 103 \\
\hphantom{$>$}2 & 16 & 11 & 31 \\
\hphantom{$>$}3 & 7 & 0 & 12 \\
\hphantom{$>$}4 & 1 & 0 & 5 \\
$>$4 & 4 & 0 & 5 \\
[4pt]
\multicolumn{1}{@{}l}{AIC} & \multicolumn{1}{c}{--} & 1\mbox{,}988.0 &1\mbox{,}481.8 \\
\hline
\end{tabular*}
\end{table}

%
%

Table \ref{distribution}
shows the distribution of the observed attacks per day compared to the
expected values under a Poisson and negative binomial probability
distribution fitted with maximum likelihood.
While the Poisson cannot capture any of the tail behavior, the negative
binomial adapts better to the tail distribution but consequently
underestimates the number of days with only~1 attack and overestimates
the number of days with 2 and 3 attacks.
A~chi-square goodness-of-fit test yields a $p$-value of $0.00016$,
suggesting that the negative binomial is not a sufficient model.
In addition, the tail is comprised of several extreme values ($36,11,10$
and 6 attacks on a single day) that exacerbate the general lack of fit.
The large number of zeros coupled with the presence of several extreme
values necessitate more complex models.


Figure \ref{figdata} also reveals the possibility that the attack days
are clustered. This is examined with Ripley's $K$-function [\citet
{Dixon2002}]. The $K$-function is a second-order function that can
reveal if clustering (or inhibition) is present in a point pattern. It
is defined loosely as
\[
K(t) \propto\bbE[\mbox{\# of future events $\leq t$ from a randomly
chosen event}].
\]
Because this measure will be effected by an inhomogeneous attack day
probability, we estimate it [\citet{baddeleyetal00}, \citet{veen06}]:
\[
\hat{K}(t) =T^{-1} \sum_{i=1}^N \frac{w_i}{\hat{p}_i}\sum_{j>i}
\frac{\mathbh{1}(t_j-t_i\leq t)}{\hat{p}_j},
\]
where $T$ is the total observational time, $\hat{p}_i$ is the estimated
probability of at least one attack at day $t_i$, and $w_i$ is a
one-sided edge correction. The estimated probability is given by the
model in (\ref{baseline}) that includes trend and seasonality
components (i.e., model BL$_5$ in Table \ref{parstrain}; see
Section \ref{results}).

\begin{figure}

\includegraphics{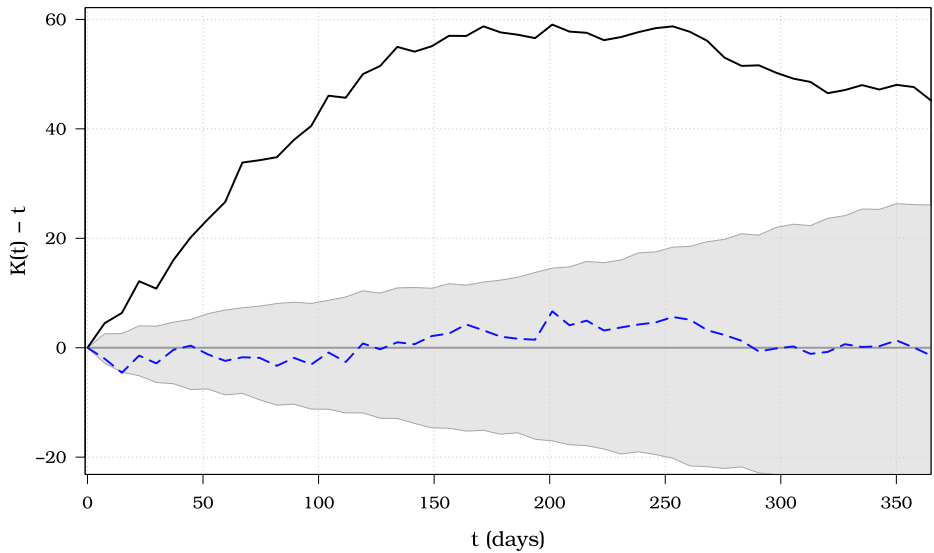}

\caption{Plot of $\hat{K}(t)-t$ for Indonesian
terrorism (1994--2000). The solid line is from the baseline model
(BL$_5$) with the gray band showing the 0.025 and 0.975 quantiles
estimated from 1,000 simulations. The dashed line is from the
self-exciting model (SE$_1$) presented in Section \protect\ref{results}.}
\label{figKplot}
\end{figure}

Figure \ref{figKplot} plots $\hat{K}(t)-t$ which has expected value 0
if the estimated probabilities are correct and there is no unaccounted
for clustering. The observed values fall well above the 95\% pointwise
confidence intervals (obtained from a parametric bootstrap, 1,000
simulations), showing that the data display clustering behavior not
accounted for by the baseline model.

%
%
%


These characteristics of the terrorism data require flexible models
that can handle the temporal attack patterns, clustering, an excess of
nonattack days and the possibility of multiple coordinated attacks on
the same day. Self-exciting hurdle models are developed in the next
section to represent the complex nature of such terrorist activity.


\section{Hurdle models and the self-exciting process}

\subsection{Hurdle models}
The hurdle models of \citet{Mullahy1986} refer to a class of two component
models for discrete count data.
These models assume that two different processes
drive the zero and nonzero counts, respectively. The \textit{hurdle}
component of
the model corresponds to the probability that the count is
nonzero
(i.e., a
terrorist action will occur), while the \textit{count} component
corresponds to
the distribution of positive counts (i.e., the number of corresponding terrorist
events). When the two components are combined, the hurdle model
produces a
probability mass function on the nonnegative integers.\vadjust{\goodbreak} Additionally,
the hurdle approach
facilitates the use of two simpler models in place of one more
complicated model.
This will be especially appealing in model fitting, as the likelihood
may be
separable, allowing for independent evaluation of each component.

For modeling the daily counts of a terrorist process, let $Y_t$ be the
number of
events on day $t$ and $E_t$ be the indicator for an \textit{event day}
such that $E_t=1$ if $Y_t\geq1$ (i.e., there is at least one terrorist
attack on day $t$)
and $E_t=0$ if $Y_t=0$ (i.e., there are no attacks on day $t$).
Also, let
$t_i$ denote the $i$th event day (not event, but day with at least one
attack).



Letting 
$\mcH_{t}=\{Y_s\dvtx s\leq t\}$ be the internal history of the
terrorist process, the hurdle component is modeled as a Bernoulli
random variable with hurdle probability
$p_t=\Pr(E_t=1|\mcH_{t-1})$ specified conditionally on the past
history of the
event process. A separate count model is constructed for
$Y_t|E_t=1$ with density $f_t(y)=\Pr(Y_t=y|E_t=1,\mcH_{t-1})$, for $y
\in\{1,2,\ldots\}$.
Because the count component is constructed conditionally on there being
at least one event, it will have no support on 0 [i.e., $f_t(0)=0$].
%
%
Combining the hurdle and count components gives the full density
%
%
\begin{equation}
f^*_t(y) = \Pr(Y_t=y|\mcH_{t-1}) = \cases{
f_t(y)p_t, &\quad$y>0$, \cr
1-p_t, &\quad$y=0$.}
\end{equation}
%

This has a similar form to a zero-inflated model [\citet
{Lambert1992}] but
differs in that for the zero-inflated model $\Pr(Y_t=0|\mcH
_{t-1})=(1-p_t)f_t(0)$.
Zero-inflated models
usually arise when there is a random censoring process that prevents
observations 
at certain times.
Alternatively, instead of assuming that terrorist events occurred but
were not
observed, the hurdle approach explicitly models the probability that no events
occurred.

The log-likelihood for a hurdle model can be decomposed as the sum of
two terms
$\log L = \log L_1 + \log L_2$, where for $T$ observations
%
%
\begin{eqnarray}
\label{bernlike}
\log L_1 &=& \sum_{t=1}^T E_t \log p_t + (1-E_t) \log(1-p_t), \\
\label{pmflike}
\log L_2 &=& \sum_{t=1}^T E_t \cdot\log f_t(y_t) \nonumber\\[-8pt]\\[-8pt]
&=& \sum_{t\dvtx E_t=1} \log f_t(y_t).
\nonumber
\end{eqnarray}
%
The first term (in the form of a Bernoulli process) represents the hurdle
component and the second term is the usual sum for a set of
observations coming
from the density $f_t$ and represents the number of events per event
day. The
form of the likelihood is equivalent to a discrete time version of a~marked point
process [\citet{DVJ03}], where the marks are
the number of attacks on an event day. A particular benefit of this
representation emerges when $f_t$ and $p_t$ share no common parameters, allowing
their log-likelihoods to be handled separately for parameter estimation.

\subsection{Self-exciting process}
It has been suggested
that some
terrorist processes exhibit self-excitation or contagion behavior
[\citet{Holden1986}, \citet{Duganetal2005}, \citet
{LaFreeetal2009}]. A
self-exciting point process [\citet{Hawkes1971a}] is one where the
realization of events increases the short-term probability of observing
future events, much in the same manner that one contagious individual
can infect other individuals (while they are still infectious) or how
major earthquakes lead to aftershocks. This type of model can be
written in the form of a cluster process [\citet{hawkesoakes1974}] and
used to explain the apparent clustering of terrorist activities.
Specifically, we consider a generalized \textit{shot noise process}
[\citet{Rice1977}], where the self-exciting component $S_t\geq0$ will
be a nonnegative function of the past history $\mcH_{t-1}$ of the form
%
%
\begin{eqnarray}
\label{selfexcite}
S_t &=& \sum_{s<t} E_s \alpha_s g(t-s) \nonumber\\[-8pt]\\[-8pt]
&=&\sum_{i\dvtx t_i < t} \alpha_i g(t-t_i) . \nonumber
\end{eqnarray}
%
The magnitude parameter, $\alpha_i$, determines the influence that
the $i$th event day has on the self-exciting process.\vadjust{\goodbreak} It 
may be a function of other associated information about the event day
(e.g., number of events, number killed, success indicator, group attribution).
Based on the results of the exploratory data analysis,
we restrict our attention to the case where
$\alpha_i\geq0$, although inhibition effects could be obtained if
negative values were
permitted.

The decay function $g(\cdot)$ specifies the shape of the excitation
based on the
time since the previous event days. To aid in parameter interpretation and
estimation, we specify $g(\cdot)$ to be a proper probability mass
function with
strictly positive support such that $g(u) \geq0$, $g(u)=0$ for $u<1$, and
$\sum_{u=1}^{\infty} g(u) = 1$. This ensures that the influence of an
event will
eventually diminish
and makes $\alpha_i$ solely
responsible for the magnitude of the effects contributed by the
$i$th event.

Any discrete distribution can be used for the decay function, provided its
support
is limited to the set of positive integers.
Standard distributions, like Poisson or negative binomial, may require shifting
or truncation to avoid having support on zero. For example, the (mean specified)
shifted negative binomial is 
%
%
\begin{equation}\label{shiftedNB}
g(u ; \mu,r) = \frac{\Gamma(r+u-1)}{\Gamma(r) (u-1)! }
\biggl(\frac{r}{\mu-1+r}\biggr)^r
\biggl(\frac{\mu-1}{\mu-1+r}\biggr)^{u-1},
\end{equation}
which has a mean of $\mu>1$, and size parameter $r>0$.
The corresponding shifted geometric distribution is given by $g(u;\mu
,r=1)$ and
the shifted Poisson density can be defined as the limiting case when
$r\rightarrow\infty$.


%
%
\begin{figure}

\includegraphics{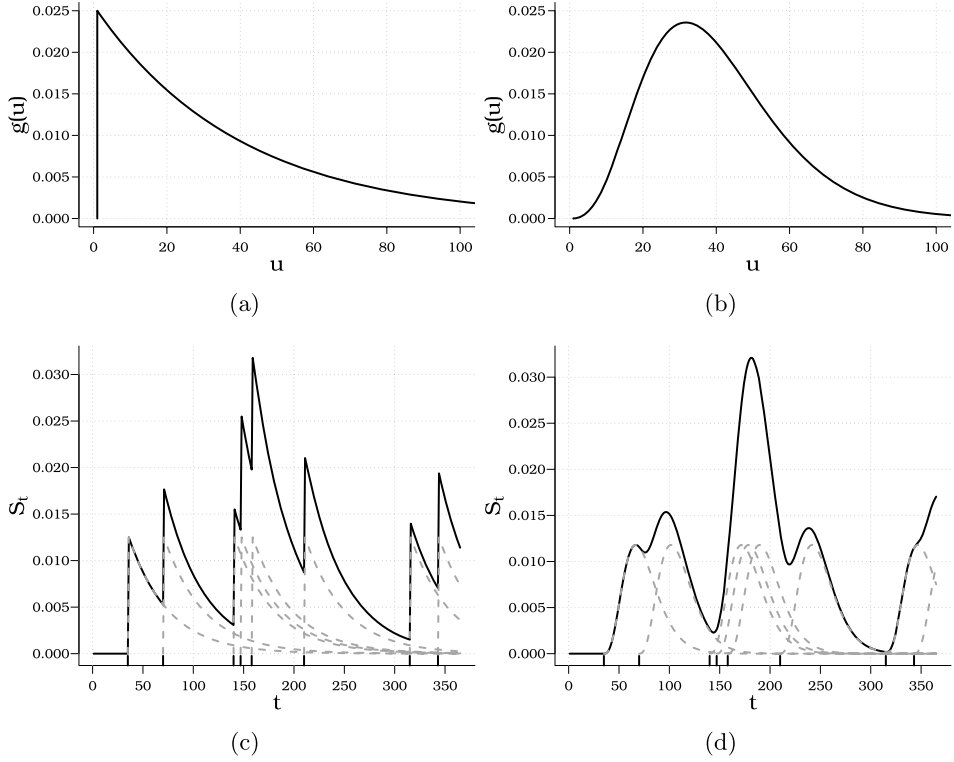}

\caption{\textup{(a)} and \textup{(b)} show the decay functions for the shifted
geometric ($\mu=40$)
and shifted negative binomial ($\mu=40,r=5$), respectively. \textup{(c)} and
\textup{(d)}
show the
corresponding shot noise processes (solid lines) and scaled component
kernels (dotted lines) for 8 events
with $\alpha_i=\alpha=0.5$.}
\label{figselfexciting}
\end{figure}
%

Figure \ref{figselfexciting} shows the geometric and negative binomial decay
functions along with the shot noise processes corresponding to 8 event days.
This illustrates how the form of the shot noise process from (\ref{selfexcite})
is similar to kernel intensity estimation
[\citet{Diggle1985}], but with weights and one-sided (predictive) kernels.

\subsection{Self-exciting Bernoulli hurdle process}
The event days, $E_t$, are modeled as a self-exciting Bernoulli process
where the probability of an event day is excited (increased) by the
occurrence of previous events. Specifically, we consider the hurdle
probability on day $t$ to be a function of the process
%
%
\begin{equation}\label{eqx}
X_t = B_t + S_t,
\end{equation}
where $B_t$ is a baseline process and $S_t$ is the self-exciting
component given
by (\ref{selfexcite}). The baseline can be a function of any exogenous
variables
that could have an effect on the process (e.g., social, political or
economic conditions and events, counter-terrorism efforts, etc.), but
it will not include any
information coming from the internal history.


To ensure the hurdle probability $p_t \in[0,1]$, a transformation
$\eta(p_t) = X_t$ is
used.
If the baseline process is nonnegative (i.e., $B_t \geq0$), then
$X_t\geq0$ and
an appealing transformation is\vadjust{\goodbreak}
$\eta(p)=-\log(1-p)$, corresponding to the hurdle probability
%
%
\begin{equation} \label{prob}
p_t=1-e^{-X_t}.
\end{equation}
%
While other transformations (e.g., logit) could be used, this one
provides several benefits.
It enjoys a likelihood function that is computationally convenient,
substituting
(\ref{prob}) into (\ref{bernlike}) obtains the Bernoulli log-likelihood
%
%
\begin{equation}\label{bernlikesimple}
\log L_1 = 
\sum_{1\leq t_i \leq T} \log(e^{X_{t_i}}-1) - \sum_{t=1}^T X_t.
\end{equation}
The shot noise component of the second sum
$(\sum_{t=1}^T X_t = \sum_{t=1}^T B_t + \sum_{t=1}^T S_t
)$ can be simplified by
recognizing 
that
\[
\sum_{t=1}^T S_t = \sum_{i\dvtx t_i<T} \alpha_i G(T-t_i),
\]
where $G$ is the cumulative density function of $g$. This significantly reduces
computation, as it is only calculated for the event days and not over
all days
in the observation window.

This transformation also allows for a straightforward interpretation of
the parameters in $X_t$.
Notice that
$p_t$ is also the probability that a Poisson random variable with rate
$X_t$ is
greater than 0.
When events are rare and~$X_t$ is small, $E_t$ can be
modeled approximately by a Poisson distribution.
According to the
superposition property of the Poisson, the event days arise from two sources,
the baseline process with rate $B_t$ and the shot noise process with rate
$S_t$. The rate from the shot noise can be further decomposed into its elements
$\alpha_ig(t-t_i)$ for $\{i\dvtx t_i<t\}$. Thus, under the Poisson
approximation,
event day $i$ will generate an expected $\sum_{u=1}^{\infty} \alpha
_i g(u) =
\alpha_i$
number of additional event days
and the decay function $g(\cdot)$ controls
when the
extra event days will occur. While these interpretations will not hold exactly
under the Bernoulli model, they do provide an indication of how each component
effects the process.

\subsection{Survival functions}
This formulation for the hurdle process also provides simple
expressions for
some properties of the time until the next event day. For a given time $t_0$,
let $\tau$ be the time of the next event day. The survival function
$V_{t_0}(u) = \Pr(\tau-t_0 > u)$
gives the probability that the next event day will be more than $u$ days
away. For the self-exciting hurdle process, this becomes
\begin{eqnarray*}
V_{t_0}(u) &=& \Pr(E_{t_0+1}=E_{t_0+2}=\cdots= E_{t_0+u}=0) \\
&=& \prod_{\delta=1}^{u} (1-p_{t_0 +
\delta}) \\
&=& \exp\Biggl\{-\sum_{\delta=1}^{u}
X_{t_0+\delta} \Biggr\}
\end{eqnarray*}
for $u \in\{1,2,\ldots\}$, where $X_{t_0 + \delta}$ is calculated
assuming no
new events have occurred since $t_0$. 
The survival function can be used to calculate the expected time until
the next
event day
\begin{eqnarray*}
\bbE_{t_0}[\tau] &=& \bbE[\tau-t_0] = 1+ \sum_{u=1}^{\infty}
V_{t_0}(u)\\
&=& 1+ \sum_{u=1}^{\infty}
\exp\Biggl\{-\sum_{\delta=1}^uX_{t_0+\delta}\Biggr\}.
\end{eqnarray*}
The survival function specification of this model can be useful for prediction,
and making inference about
future attacks given the current history.
\section{Results}
\label{results}
In order to build the models and evaluate their predictive performance,
the data
are partitioned into two time periods.
The first time period from 1994 through 2000 is used to construct the
models and
estimate their parameters.
The second period from 2001 through 2007 is used to assess the predictive
performance of the models.


\subsection{Event day modeling}

Recall that the hurdle probability is construc\-ted from the sum of a
baseline and self-exciting
processes (\ref{eqx}).
In order to capture potential seasonality and other large scale trends
in the data, the baseline process $B_t$ is defined as
%
%
\begin{equation}\label{baseline}
\log(B_t) = \beta_0 +\beta_1 t + \beta_2 t^2 + A_1 \sin(2\pi t/
\omega) + A_2
\cos(2\pi t /\omega),
\end{equation}
where the seasonal terms have a period of $\omega=365.25$ days.
This results in five baseline parameters
$\theta_B = (\beta_0,\beta_1,\beta_2,A_1,A_2) \in\bbR^5$. The log
transform
ensures $B_t\geq0$.
We considered self-exciting processes of the form 
%
%
\begin{equation}\label{selfexcite2}
S_t(\alpha,\mu,r) = \alpha\sum_{i\dvtx t_i<t} g(t-t_i;\mu,r),
\end{equation}
where $g(u;\mu,r)$ is a shifted negative binomial decay function
(\ref{shiftedNB}) and the magnitude $\alpha$ is a constant.
This results in three parameters for the self-exciting component
$\theta_S = (\alpha,\mu,r) \in\bbR_+^3$.

The combined model for $X_t=B_t+S_t$ 
results in up to
8 parameters to estimate. Estimation is carried out by maximizing the
log-likelihood function given in (\ref{bernlikesimple}). Since there
is no
explicit solution available, we used \textsf{R}'s numerical
optimization routine \texttt{nlminb} [\citet{R}] to obtain estimates.
For all models, the estimates converged
within seconds from a variety of starting points.

%
%
\begin{table}
\caption{Parameter estimates for the event day models for the training
period (1994--2000). The BL are the baseline only and the SE are the
self-exciting models}
\label{parstrain}
\begin{tabular*}{\tablewidth}{@{\extracolsep{\fill}}lcd{2.2}d{3.2}d{2.2}d{2.2}d{2.2}ccc@{}}
\hline
\textbf{Model} & \multicolumn{1}{c}{\textbf{AIC}} & \multicolumn{1}{c}{$\bolds{\beta_0}$} & \multicolumn{1}{c}{$\bolds{\beta_1}$}
& \multicolumn{1}{c}{$\bolds{\beta_2}$} & \multicolumn{1}{c}{$\bolds{A_1}$} & \multicolumn{1}{c}{$\bolds{A_2}$} &
\multicolumn{1}{c}{$\bolds{\alpha}$} & \multicolumn{1}{c}{$\bolds{\mu}$} & \multicolumn{1}{c@{}}{$\bolds{r}$} \\
\hline
BL$_1$ & 1,187.77 & -2.75 & & & & & & & \\
BL$_2$ & 1,149.54 & -8.2 & 8.05 & & & & & & \\
BL$_3$ & 1,151.25 & -2.83 & -8.01 & 11.91 & & & & & \\
BL$_4$ & 1,171.22 & -2.82 & & & -0.42 & -0.31 & & & \\
BL$_5$ & 1,136.80 & -8.02 & 7.69 & & -0.35 & -0.32 & & & \\
BL$_6$ & 1,138.60 & -3.53 & -5.74 & 9.95 & -0.35 & -0.32 & & & \\
\textbf{SE$_1$} & \textbf{1,075.21} & \multicolumn{1}{c}{\textbf{$\bolds{-}$4.41}\hspace*{2.5pt}} & & & & &
\textbf{0.89 }& \textbf{37.54} & \textbf{0.45} \\
SE$_2$ & 1,077.16 & -5.17 & 1.23 & & & & 0.87 & 36.55 & 0.45 \\
SE$_3$ & 1,078.71 & 15.31 & -63.2 & 50.24 & & & 0.85 & 35.69 & 0.45 \\
SE$_4$ & 1,077.03 & -4.34 & & & -0.41 & -0.40 & 0.86 & 38.55 & 0.44 \\
SE$_5$ & 1,078.76 & -5.99 & 2.68 & & -0.38 & -0.42 & 0.82 & 36.57 & 0.43
\\
SE$_6$ & 1,079.60 & 20.46 & -79.33 & 62.91 & -0.43 & -0.45 & 0.80 &
37.28 & 0.43 \\
\hline
\end{tabular*}
\end{table}

\begin{figure}[b]

\includegraphics{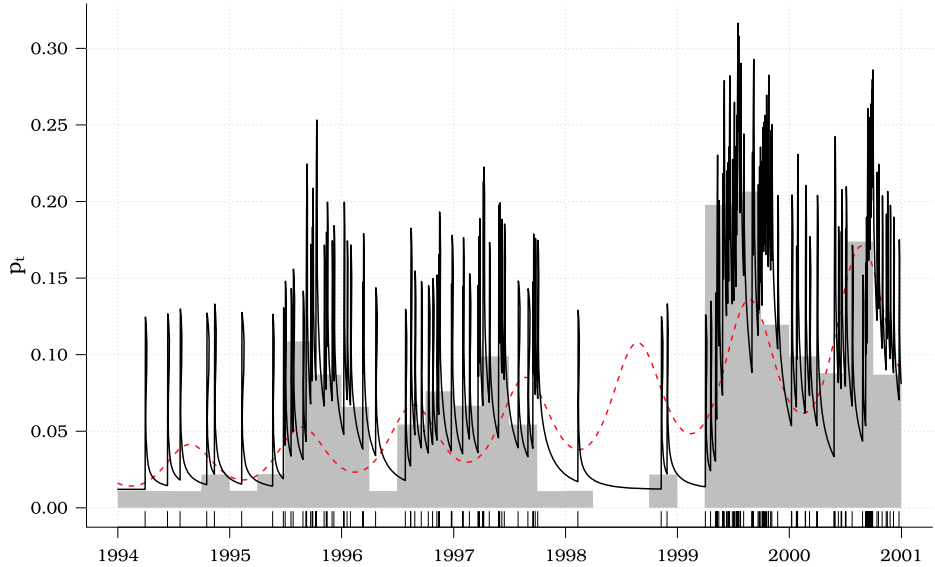}

\caption{Hurdle probabilities for the best fitting self-exciting model
SE$_1$ (solid line), the best fitting baseline only model BL$_5$
(dotted line), and the observed piecewise constant probabilities
(blocks). The rug signifies the event days.}
\label{fighurdleProbtraining}
\end{figure}

Table \ref{parstrain} shows the evaluated models and their
corresponding AIC
scores. The lowest AIC belongs to the four component model with a constant
baseline (SE$_1$), providing essentially a discrete time version of the
Hawkes model
[\citet{Hawkes1971a}, \citet{ozaki79}].
This AIC is much lower than the best baseline only model (BL$_5$) which
includes the periodic terms.
Figure \ref{figKplot} plots the weighted
$K$-function for SE$_1$ and BL$_5$ showing the self-exciting model is
much tighter around 0 indicating a better fit over the baseline only
model [\citet{Diggle79}].
Figure \ref{fighurdleProbtraining} shows the estimated hurdle
probability in the training period for both models.

%

The parameters of the self-exciting model (SE$_1$) suggest a baseline
hurdle probability of about
$0.012$ 
when the shot noise term drops to zero.
However, according to the Poisson approximation, every event day will
generate an expected $\alpha=0.89$ additional event days. The
generated event day is expected to occur in $\mu=37.54$ days, but the
time until occurrence has a median of only 16 days.

\subsection{Count modeling}




It can be obtained from Table \ref{distribution} that most event days
(92.4\%) are comprised of only 1 or 2
attacks. However, several days (1.9\%) have more than 9 attacks (with
one day
having 36 recorded terrorist attacks).
This suggests count distributions that have an initial rapid decay, but still
possess a long tail to accommodate the extreme values.
One such discrete distribution 
is
the Riemann zeta or discrete Pareto distribution.
This power law distribution
has recently been employed to model the severity (e.g., number killed
or injured) of terrorist attacks [\citet{clauset2007}].

Using it to model the event day
counts, under an i.i.d assumption, gives the probability mass function
\[
f_t(y;s) = \frac{y^{-s}}{\zeta(s)},\qquad  y \in\{1,2,\ldots\},
\]
where $\zeta(s)=\sum_{x=1}^{\infty} x^{-s}$ is the Riemann zeta
function and the parameter $s\in(1,\infty)$.
This one parameter model is easy to estimate numerically via maximum
likelihood [\citet{goldstein04}, \citet{seal52}] 
by finding the $s$ that satisfies the equation
$\frac{1}{n}\sum_{i=1}^n \log(y_i) = -\frac{\zeta'(s)}{\zeta(s)}$.
The training period data gives rise to an estimated parameter of $\hat
{s}=2.86$.
The bootstrap Kolmogorov--Smirnov test of \citet{clauset09} gives a~$p$-value of $0.69$,
suggesting the zeta distribution provides a suitable fit.
%

While the constant parameter zeta distribution appears to provide a
good fit,
there may be better models, especially if the distribution is allowed
to vary
over time.
In particular, as for the hurdle component, distributions that
vary over time in response to a shot noise process are considered. 
An additional driving process (with shot noise) was implemented for the
count model, namely,
%
\[
X^c_t(\beta^c,\alpha^c,\mu^c) = \beta^c + \alpha^c\sum_{i\dvtx t_i<t}
Y_i \cdot g^c(t-t_i;\mu^c),
\]
%
where $\beta^c$ is a constant baseline, $\alpha^c_i=\alpha^c\cdot
Y_i$ varies in response to the number of attacks on the $i$th event
day, and the decay function $g^c(u;\mu^c)$ is the one parameter
shifted geometric function with mean $\mu^c$. This results in three
parameters for the count component $\theta_C=(\beta^c,\alpha^c,\mu^c)$.
Note that this is a~separate process than the one specified for the
hurdle component given by~(\ref{baseline}) and (\ref{selfexcite2}).
The additional flexibility afforded by the hurdle specification allows
such additional complexity to be introduced for the count component of
the model.

There are two primary hypotheses about how the count distribution
behaves in response to the event history. First, counts could respond
in a~self-exciting manner, where recent event and multi-event days
increase the likelihood of further multi-event days.
Alternatively, due to the depletion of resources or increased
anti-terrorism measures after large attacks, counts may respond in a
more self-inhibiting manner, where recent multi-event days lower the
chances of multi-events day in the near future.

%
\begin{table}
\caption{Parameter estimates for the count models for the training
period (1994--2000). C$_z$, C$_{\mathit{se}}$ and C$_{\mathit{si}}$ are the constant
parameter, self-exciting and self-inhibiting forms, respectively, of
the Riemann zeta distribution}
\label{countpars}
\begin{tabular*}{\tablewidth}{@{\extracolsep{\fill}}lccd{1.3}cd{2.2}@{}}
\hline
\textbf{Model} & \multicolumn{1}{c}{\textbf{AIC}} & \multicolumn{1}{c}{$\bolds{s}$} & \multicolumn{1}{c}{$\bolds{\beta}$}
& \multicolumn{1}{c}{$\bolds{\alpha}$} & \multicolumn{1}{c@{}}{$\bolds{\mu}$} \\
\hline
C$_z$ & 241.00 & 2.86 & & & \\
C$_{\mathit{se}}$ & 240.48 & $(1-e^{-X_t})^{-1} $& 0.375 & 0.20 & 2.13 \\
C$_{\mathit{si}}$ & 244.72 & $ e^{X_t}$ & 1.00 & 0.34 & 55.18 \\
\hline
\end{tabular*}
\vspace*{-3pt}
\end{table}

%
\begin{figure}[b]
\vspace*{-3pt}
\includegraphics{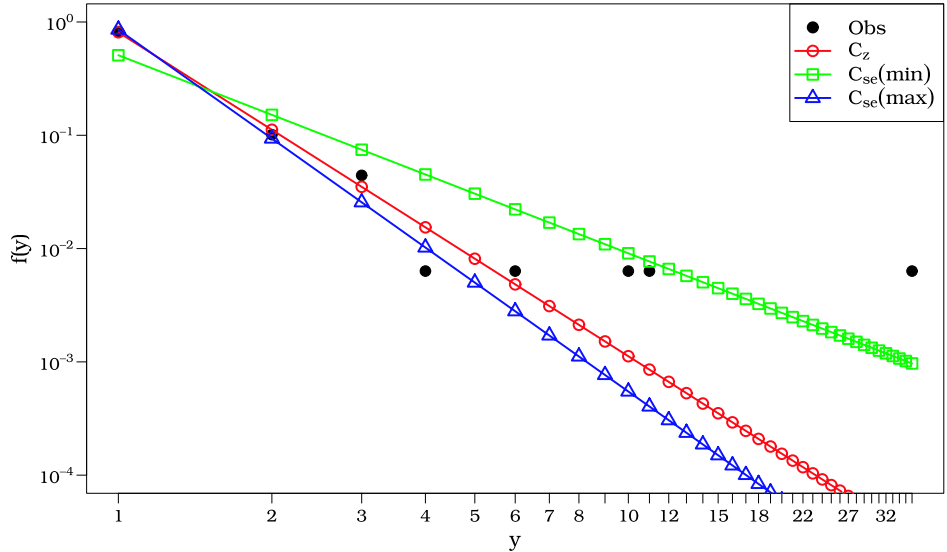}

\caption{Estimated count probabilities for the constant parameter
model (C$_z$ with $s=2.86$) and for the self-exciting count model
(C$_{\mathit{se}}$) at the baseline state ($s=3.19$) and most excited state ($s=1.75$).}
\label{figcountdensity}
\end{figure}

To evaluate the self-exciting hypothesis, let $s_t=(1-e^{-X_t^c})^{-1}$
since smaller values of $s$ lead to more mass on the extreme values.
Alternatively, $s_t=e^{X_t^c}$ for the self-inhibiting model.
Table \ref{countpars} shows the estimated parameters and AIC scores
for the fitted count models. The lower AIC suggests that the
self-exciting count model has more validity than the self-inhibiting
model for the training data.
Figure~\ref{figcountdensity} shows the estimated probability of counts
under the constant parameter (C$_{z}$) and self-exciting (C$_{\mathit{se}}$) models.
While the self-exciting model puts more mass on multi-event days when
$X_t^c$ is large, the
decay is rapid and the self-exciting behavior is short lived---only for
a few days, before the shot noise term approaches 0 and its effects become
negligible.

\subsection{Testing model predictions}
Based on the training period, models are selected and fit
for both the hurdle and count components.
To evaluate\vadjust{\goodbreak} how well this can predict future terrorist activity, 
the models are applied to the terrorist incidents in the testing
period (2001--2007) using the following procedure:
\begin{longlist}[(1)]
\item[(1)] At day $t$ predict the probability of an event day and
conditional number of
attacks on day $t+1$, based on the current history $\mcH_{t}$.
\item[(2)] Observe $Y_{t+1}$ 
and
re-estimate the model parameters based on the updated history
$\mcH_{t + 1}$.
\item[(3)] Repeat steps (1) and (2) through the
testing period to construct a~final model based on all data between
1994--2007.
\end{longlist}
%
%
%
This update and forecast scheme results in a set of daily predictions
$\hat{p}_t = \Pr(E_t=1|\mcH_{t-1})$ and 
$\hat{f}_t(y) = \Pr(Y_t=y|E_t=1,\mcH_{t-1})$
conditioned on the past history of the process.

The predictive capability of the models are evaluated by comparing the
log probability gain, or predictive log-likelihood ratio, of each model
[\citet{daleyVJ04}]. This score is a relative measure of the
improvement in predicative capability for a model compared to a
reference or null model. The interpretation is that values greater than
$0$ show a relative improvement in predictions over the reference
model, with the highest values indicating the preferred model.

%
%
\begin{table}[b]
\caption{The hurdle component log probability score, $G$, using the
constant baseline only model (BL$_1$) as the reference model. The AIC
scores are from the training period (see Table \protect\ref{parstrain})}
\label{infogain}
%
%
\begin{tabular*}{\tablewidth}{@{\extracolsep{\fill}}lcd{3.2}ccc@{}}
\hline
\textbf{Model} & \multicolumn{1}{c}{\textbf{AIC}} & \multicolumn{1}{c}{$\bolds{G}$} & \multicolumn{1}{c}{\textbf{Model}}
& \multicolumn{1}{c}{\textbf{AIC}} & \multicolumn{1}{c@{}}{$\bolds{G}$} \\
\hline
BL$_1$ & 1,187.77 & 0.00 & \textbf{SE$_1$} & \textbf{1,075.21} &
\textbf{26.96}\\
BL$_2$ & 1,149.54 & -19.10 & SE$_2$ & 1,077.16 & 25.47 \\
BL$_3$ & 1,151.25 & 20.69 & SE$_3$ & 1,078.71 & 24.96 \\
BL$_4$ & 1,171.22 & -1.39 & SE$_4$ & 1,077.03 & 26.22 \\
BL$_5$ & 1,136.80 & -18.21 & SE$_5$ & 1,078.76 & 24.56 \\
BL$_6$ & 1,138.60 & 21.63 & SE$_6$ & 1,079.60 & 24.57 \\
\hline
\end{tabular*}
\end{table}

For the hurdle component, the log probability gain is
\[
G = \sum_{t \in\mcT_2} E_t \log\frac{\hat{p}_t}{\hat\pi_t} +
(1-E_t) \log\frac{1-\hat{p}_t}{1-\hat\pi_t},
\]
where $\hat{p}_t$ comes from the model of interest and $\hat\pi_t$
from the constant baseline only model (BL$_1$), and $\mcT_2=[01$ Jan
2001, 31 Dec 2007] covers the testing period.
Table \ref{infogain} compares the log probability gains $G$ and the
AIC scores from the training period, and shows that the results from
the test period were similar to those in the training period,
supporting the use of AIC for model selection.

%

The constant baseline self-exciting model (SE$_1$) has the largest log
probability
gain and all self-exciting models outperform all of the baseline only models.
The best baseline only model for the testing period was the
full model with seasonality and a quadratic trend (BL$_6$).
This is in contrast to the training period results where the best
baseline only model had a linear trend (BL$_5$). The linear trend model
failed to
perform as well during the testing period, as the general attack rate
peaked around
the transition point and started decreasing during the remainder of the
testing period, creating a significant quadratic trend over the
combined observation
period. Alternatively, the self-exciting models can adapt rapidly to
sudden changes
in attack rate, yielding better overall predictive ability.

The count component uses the log probability gain score
\[
G^c = \sum_{i\dvtx t_i \in\mcT_2} \log\frac{\hat{f}(y_i)}{\hat{h}(y_i)},
\]
where $\hat{f}$ is the model of interest and $\hat{h}$ is the
reference model.
Table \ref{Gcounts} shows the values of $G^c$ using the constant parameter
zeta model (C$_z$) as the reference model. These results show the self-exciting
zeta model (C$_{\mathit{se}}$) as the preferred model for predicting the counts.
%
%
%
\begin{table}
\tablewidth=190pt
\caption{The count component log probability score, $G^c$, using the
constant parameter zeta model (C$_z$) as the reference model}
\label{Gcounts}
\begin{tabular*}{\tablewidth}{@{\extracolsep{\fill}}lrrr@{}}
\hline
& \textbf{C$\bolds{_s}$} & \textbf{C$\bolds{_{\mathit{se}}}$} & \textbf{C$\bolds{_{\mathit{si}}}$} \\
\hline
$G^c$ & 0.00 & 8.07 & $-$0.27 \\
\hline
\end{tabular*}
\end{table}
As with the hurdle component, the predictive performance of the count
models followed the AIC scores from the training period. However, the
model with the self-exciting component does predict substantially
better than the constant component model---more so than would be
expected from the AIC scores in the training period (see Table~\ref
{countpars}). This may be due to changes in the attack behavior during
the testing period.
Figure \ref{figdata} shows that while there were a few large counts
during 2001, there were very few multi-attack days after.
This change in the pattern of multiple attacks accounts for the
improved predictive capabilities of the self-exciting model over the
constant parameter model. The self-exciting count model is able to
quickly adapt to changing conditions and consequently makes better predictions.


In summary, the daily counts of Indonesian terrorist attacks were
modeled with a self-exciting hurdle model. The separation of the model
into two components allowed both the attack rate and attack
characteristics to be more faithfully represented.
It was found that including a self-exciting process benefited both the
hurdle and count components by allowing the models to quickly adapt to
changing attack behavior.
It is this property as a predictive tool that makes the self-exciting
hurdle model especially useful in an applied setting where prediction
is of primary interest.



\section{Conclusion}
%
%
In practice, policy makers are
faced with limited resources, financially, materially and in personnel. The
allocation of these resources depends on an understanding of the future
risk of
terrorist activity.
As an example, in the wake of the attacks of September 11, 2001, policy
makers placed additional security in the form of National Guard troops
in all US commercial airports. This was done in hopes of
re-establishing public trust in the safety of air travel while the
intelligence community was assessing the risk of subsequent attacks
[\citet{PriceForrest2009}].
The obvious question in that case was when to shift
resources from increased airport security to investing in long-term strategies
to improve intelligence capabilities.
As this case illustrates, only by clearly understanding the risk of
terrorist activity can policy makers make informed decisions about
allocating resources.

This paper presents a two-component self-exciting model for the
analysis and prediction of future
terrorist activity.
It was found that the best model used a constant baseline with a
self-exciting term for the hurdle component and a Riemann zeta
distribution with shot noise driven parameter for the count component.
The improvement offered by the self-exciting term provides support for
the contagion theory and suggests a significant short term increase in
terrorism risk after an attack. The use of a power law distribution for
the number of same day attacks corresponds with the power law behavior
of the number killed in terrorist attacks [\citet{clauset2007}]. This
is to be expected, as multiple coordinated attacks would tend to
indicate better planned, and hence more severe, terrorist activities.

The self-exciting hurdle model adheres to the theoretical concept for a~contagion effect to terrorism as manifested in the clustering of data while
providing good fit and predictive capabilities without relying on
exogenous variables.
The model provides a simple structure and interpretation of the
parameters useful for understanding the dynamic nature of the terrorist
activity.
This provides an appropriate staring point for exploring
additional covariate effects, including analysis concerning the effectiveness
of counter-terrorism activities, geography, political or economical factors.
As the attack clustering may be partially attributable to a stochastic
baseline driven by such exogenous processes [\citet{Holden1986}],
further analysis could include covariates in the baseline model (\ref
{baseline}) to test the impact on the self-exciting component.

By focusing on the timing of terrorist attacks on the islands of
Indonesia, other aspects of the attacks, such as location, attack type
and group responsible, were not considered. Including such additional
information could lead to more detailed models that better explain the
nuances of the terrorist activity. For example, the self-exciting
component could include a~spatial proximity term using the models of
\citet{Mohleretal2010} or multiple self-exciting terms could be
included that represent how attacks from one terrorist group influence
the subsequent attacks of other groups.


The utility of the models presented here and their ease of
implementation and
interpretation make them a potentially useful tool in security related
fields. The
results show that the risk of terrorist activity can vary greatly over short
periods of time, thus policy responses in terms of resource allocation, security
and counter-terrorism responses should reflect this as well as
addressing the
more long-term trends in risk.
For example, understanding the short-term
variations in risk would allow a more effective deployment and
assessment of
additional airport
security measures in the wake of a terrorist attack, while a grasp of
the longer
term trends could help guide the development and assessment of more strategic
counter-terrorism
resources, such as increasing the number of foreign language experts available
for translation or developing effective de-radicalization programs to
prevent the
growth of terrorist groups.

\printaddresses

\end{document}